\documentclass[a4paper]{PoS}
\usepackage{amsfonts}
\usepackage{amsthm}
\usepackage{amssymb,amsmath}
\usepackage{mathtools}
\usepackage{hepunits}
\newcommand*\diff{\mathop{}\!\mathrm{d}}

\usepackage{braket}
\usepackage[dvipsnames]{xcolor}
\usepackage{graphicx}
\graphicspath{ {image/} }
\usepackage{subfigure} 
\usepackage{multirow}
\title{Frame dependence of transition form factors in light-front dynamics}

\ShortTitle{Frame dependence of transition form factors in light-front dynamics}

\author{\speaker{Meijian Li}\\
        Iowa State University\\
        E-mail: \email{meijianl@iastate.edu}}

\abstract{
We study the radiative transitions between vector and pseudoscalar quarkonia in the light-front Hamiltonian approach, and investigate the effects of using different current component and different reference frames. In practical calculations with truncated Fock spaces, transition form factors may acquire current component dependence and frame dependence, and such dependences could serve as a measure for the Lorentz symmetry violation. We suggest using the transverse current with $m_j=0$ state of the vector meson, since this procedure employs the dominant spin components of the light-front wavefunctions and is more robust in practical calculations. 
We calculate the transition form factor between vector and pseudoscalar quarkonia 
and investigate the frame dependence with light-front wavefunctions calculated from the valence Fock sector. We suggest using frames with minimal longitudinal momentum transfer for calculations in the valence Fock sector, namely the Drell-Yan frame for the space-like region and a specific longitudinal frame for the timelike region; at $q^2=0$ these two frames give the same result.
}

\FullConference{Light Cone 2019 - QCD on the light cone: from hadrons to heavy ions - LC2019\\
		16-20 September 2019\\
		Ecole Polytechnique, Palaiseau, France}

\begin{document}


\section{Transition form factor}
The transition between a pseudoscalar and a vector meson via emission of a photon, $\mathcal{V}\to \mathcal{P} \gamma$, also known as the magnetic dipole (M1) transition, offers insights into the internal structure of the meson bound states.
The underlying dynamics is encoded within the transition form factor $V(q^2)$, which arises from the Lorentz structure decomposition of the hadron matrix element as~\cite{Li:2018uif,Li:2019kpr}
 \begin{align}\label{eq:Vq2_def}
   I^\mu_{m_j}\equiv  \bra{\mathcal{P}(P)} J^\mu(0) \ket{\mathcal{V}(P',m_j)}
   =\frac{2 V(q^2)}{m_{\mathcal{P}}+m_{\mathcal{V}}}\epsilon^{\mu\alpha\beta\sigma}{P}_\alpha P'_\beta e_{\sigma}(P', m_j)
   \;,
 \end{align}
where $q^\mu = {P'}^\mu - P^\mu$. $m_{\mathcal{P}}$ and $m_{\mathcal{V}}$ are the masses of the pseudoscalar and the vector, respectively. 
 $e_{\sigma}$ is the polarization vector of the vector meson, and $m_j=0,\pm 1$ is the magnetic projection.  

 On the light front, $V(q^2)$ cn be extracted from any of the four current components, $\mu= +,-,x,y$, and with different magnetic projections of the vector meson. We summarize these formulas 
 in Table.~\ref{tab:V_component_mj}. Note that we use $J^{R/L} \equiv J^x \pm i J^y$ as the transverse currents.
 We introduce two variables, $z\equiv ({P'}^+-P^+)/{P'}^+$ and $\vec \Delta_\perp\equiv\vec q_\perp-z\vec P'_\perp$ such that $q^2=zm_{\mathcal{V}}^2-m_{\mathcal{P}}^2 z/(1-z)-\vec\Delta_\perp^2 /(1-z)$~\cite{Li:2019kpr,Li:2017uug}. 
 For each possible value of $q^2$, the different choices on the value of the pair $(z,\vec\Delta_\perp)$ correspond to different frames. In particular, $z=0$ corresponds to the Drell-Yan frame and we define $\vec \Delta_\perp=0$ as the longitudinal frame. The latter has two branches, $z=[m_{\mathcal{V}}^2-m_{\mathcal{P}}^2+q^2\pm\sqrt{(m_{\mathcal{V}}^2-m_{\mathcal{P}}^2+q^2)^2-4m_{\mathcal{V}}^2q^2}]/(2m_{\mathcal{V}}^2)$, namely longitudinal-I/II.

\begin{table}[b]
  \centering 
   \caption{The formulas of extracting the transition form factor $V(q^2)$ from different current components and different $m_j$ states of the vector meson, derived from Eq.~\eqref{eq:Vq2_def}. 
   }\label{tab:V_component_mj}
   \resizebox{\textwidth}{!}
   {
   \begin{tabular}{c ccc}
     \hline\hline\\[-4mm]
     $\dfrac{2 V(q^2)}{m_{\mathcal{P}}+m_{\mathcal{V}}}$ 
     &$m_j = 0$
     &$m_j = 1$
     &$m_j = -1$\\[-4mm]\\
     \hline\\[-3mm]
     $J^+$    
& -
&\(\dfrac{i\sqrt{2}I^+_1}{{P'}^+\Delta^R}\)
&\(\dfrac{-i\sqrt{2}I^+_{-1}}{{P'}^+\Delta^L}\)
\\[1pt]\\
     $J^R$    
&  \(\dfrac{-i I^R_0}{m_{\mathcal{V}}\Delta^R}\)
&\(\dfrac{i\sqrt{2}I^R_{1}}{{P'}^R\Delta^R}\)
&\(~~~\dfrac{i\sqrt{2}(1-z)I^R_{-1}}{(m_{\mathcal{P}}^2 - (1-z)^2 m_{\mathcal{V}}^2-P^R\Delta^L )}~~~\)
     \\[1mm]\\
     $J^L$
&\(\dfrac{iI^L_0}{m_{\mathcal{V}}\Delta^L}\)
&\(\dfrac{-i\sqrt{2}(1-z)I^L_{1}}{(m_{\mathcal{P}}^2 - (1-z)^2 m_{\mathcal{V}}^2-P^L\Delta^R )}~~~\)
&\(\dfrac{-i\sqrt{2}I^L_{-1}}{{P'}^L\Delta^L}\)
     \\[1mm]
     \\
$J^-$
&\(\dfrac{-iP^+ I^-_0}{m_{\mathcal{V}}( \Delta^R P^L-\Delta^LP^R )}\)
&\(\dfrac{-i\sqrt{2}P^+ {P'}^+I^-_{1}}{{P'}^+{P'}^R(m_{\mathcal{P}}^2 - P^L\Delta^R )- P^+P^R m_{\mathcal{V}}^2}\)
&\( \dfrac{i\sqrt{2}P^+ {P'}^+I^-_{-1}}{{P'}^+{P'}^L(m_{\mathcal{P}}^2 - P^R\Delta^L )- P^+P^L m_{\mathcal{V}}^2}\)
\\[-2mm]
\\
\hline\hline
   \end{tabular}
   }
 \end{table}
\section{In the valence Fock sector}
Though in principle, the transition form factor $V(q^2)$ is Lorentz invariant, practical calculations with truncated Fock space could break the Lorentz symmetry, and spurious dependences on the current component and reference frame could emerge.
In the valence Fock sector, we take the impulse approximation
and calculate the hadron matrix element for the quark as Eq.~\eqref{eq:jmuq}. We therefore compute $\hat{V}(q^2)$ which is related to $V(q^2)$ as \( V(q^2)=2e \mathcal{Q}_f \hat{V}(q^2) \). $\mathcal{Q}_f$ is the dimensionless fractional charge of the quark.
\begin{align}\label{eq:jmuq}
  \begin{split}
    I^\mu_{q,m_j}\equiv  &
  \bra{\mathcal{P}(P)} J_q^\mu(0) \ket{\mathcal{V}(P',m_j)}
    =
    \sum_{s,\bar s}
    \int_z^1\frac{\diff x'}{2x'(1-x')}
    \int\frac{\diff^2 k'_\perp}{{(2\pi)}^3}
    \frac{1}{x}\\
    &\times
    \sum_{s'}
    \psi_{s' \bar s/\mathcal{V}}^{(m_j)}(\vec k'_\perp, x')
    \psi_{s \bar s/\mathcal{P}}^{*}(\vec k_\perp, x)  
     \bar{u}_{s  }(x{P}^+ , \vec k_\perp+x\vec P_\perp)
    \gamma^\mu u_{s'}(x'{P'}^+, \vec k'_\perp +x'\vec P'_\perp)
    \;.
  \end{split}
\end{align}
$\psi^{(m_j)}_{s\bar s/h}(\vec k_\perp, x)$ is the light-front wavefunction (LFWF) written in relative coordinates $x\equiv p^+/P^+$ and $\vec k_\perp\equiv \vec p_\perp-x\vec P_\perp$, where $p$ is the single-particle 4-momentum of the quark. $s$ represents the fermion spin projection. The initial and final states are related as $x'=x+z(1-x)$ and $\vec k_\perp'=\vec k_\perp + (1-x)\vec \Delta_\perp$.

\subsection{The preferred current}
In the valence Fock sector, the transition form factor is calcualted by applying Eq.~\eqref{eq:jmuq} to the formulas in Table~\ref{tab:V_component_mj}. We find that only two sets of those options 
could unambiguously extract the transition form factor: $\hat V|_{J^{R/L}, m_j=0}$ 
and $\hat V|_{J^+, m_j=\pm 1}$ 
. 
For all other choices, fixing the values of $z$ and $\Delta_\perp$ could not uniquely determine the transition form factor, since there is always an extra dependence on the transverse momentum of the meson, $\vec P_\perp$, or equivalently on $\vec {P'}_\perp$. This implies that the resulting transition form factors are not invariant under the transverse boost. 

For the two possible choices,
$\hat V|_{J^{R/L}, m_j=0}$ mainly employs the overlap of the dominant components, $\psi_{\uparrow\downarrow+\downarrow\uparrow/\mathcal{V}}^{(m_j=0)}\psi_{\uparrow\downarrow-\downarrow\uparrow/\mathcal{P}}^{ *}$, whereas even the major part of $\hat V|_{J^+, m_j=\pm 1}$ involves the subdominant components, such as $\psi_{\uparrow\uparrow/\mathcal{V}}^{(m_j=1)}\psi_{\uparrow \uparrow/\mathcal{P}}^{ *}$.
This suggests that $\hat V|_{J^{R/L}, m_j=0}$ is more reliable for practical calculations where the dominant components of a system are better constrained than the subdominant ones~\cite{Li:2018uif}. 

We adopt wavefunctions of heavy quarkonia and light mesons from the Basis Light-Front Quantization (BLFQ) approach~\cite{1stBLFQ, Yang_run, WQ_light}. The effective Hamiltonian extends the holographic QCD~\cite{holography} by introducing the one-gluon exchange interaction with a running coupling. 
In Table~\ref{tab:proportion}, we list the proportions of the dominant and subdominant spin components for the low lying vector and pseudoscalar states. The dominant terms are profound even in the light mesons.
\begin{table}[h]
  \caption{
    \label{tab:proportion}
    Proportions of the dominant and subdominant spin components for the low lying states of the heavy quarkonia and light mesons~\cite{Yang_run, WQ_light}.
  }
  \centering
  \begin{tabular}{l |ccc| ccc}
    \hline
    \hline
  \multirow{2}{*}{$\%$} 
  & \multicolumn{3}{c }{$\mathcal{V} [m_j=0, m_j=1]$} 
  & \multicolumn{3}{|c }{$\mathcal{P}$} \\
  \cline{2-7}
  &$\Upsilon$ & $J/\psi$ & $\rho$
  &$\eta_b$ & $\eta_c$ & $\pi$\\
  \hline
   dominant
  &100, 98.5 & 99.9, 95.9 & 98.8, 84.2
  &96.6 & 88.0 & 62.9\\
   subdominant
  &0, 1.5 & 0.1, 4.1 & 1.2, 15.8
  &3.4 & 12.0 & 37.1\\
  \hline  \hline
  \end{tabular}
  \end{table}

\begin{figure}[t]
  \centering
  \includegraphics[width=0.3\textwidth]{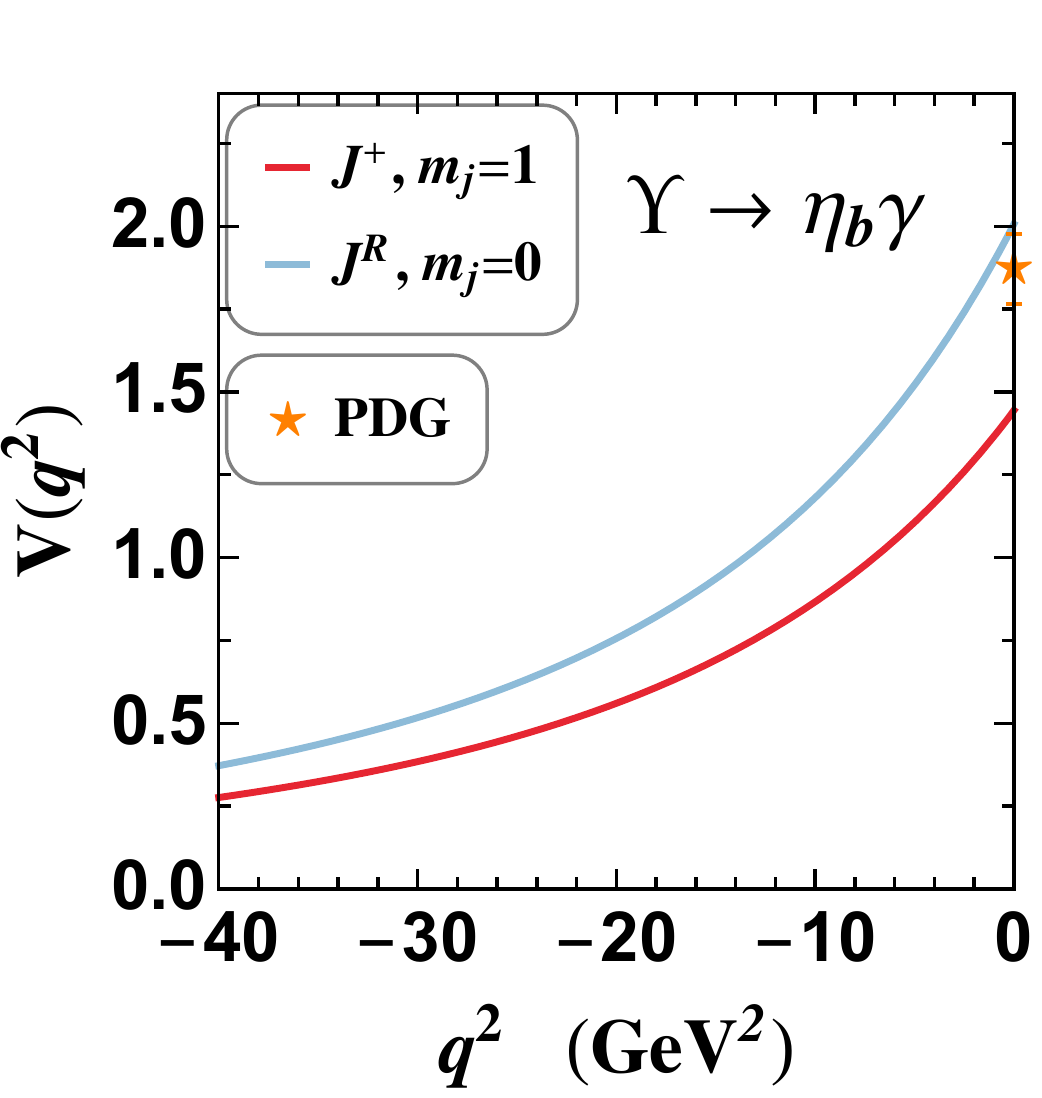}
  \includegraphics[width=0.3\textwidth]{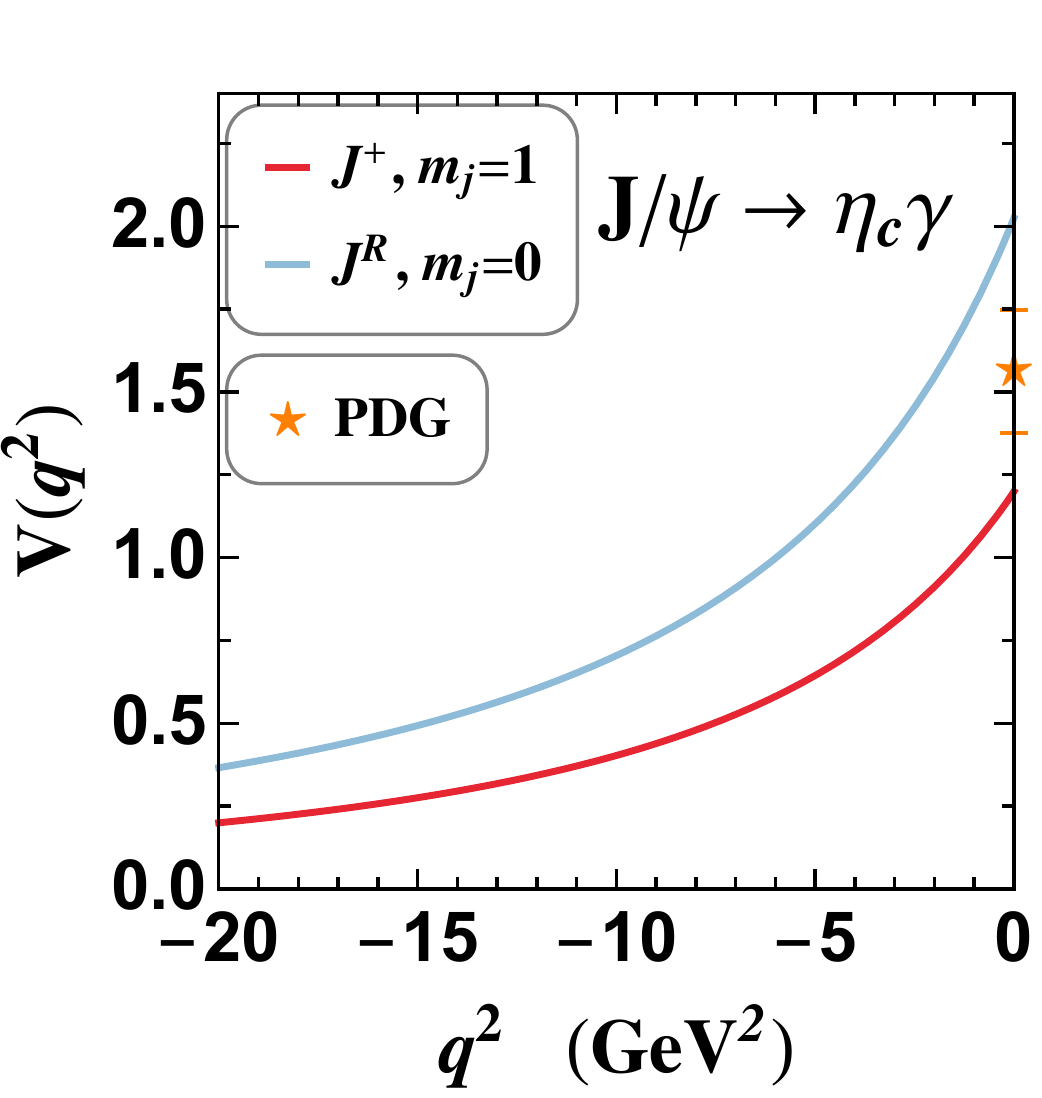}
  \includegraphics[width=0.3\textwidth]{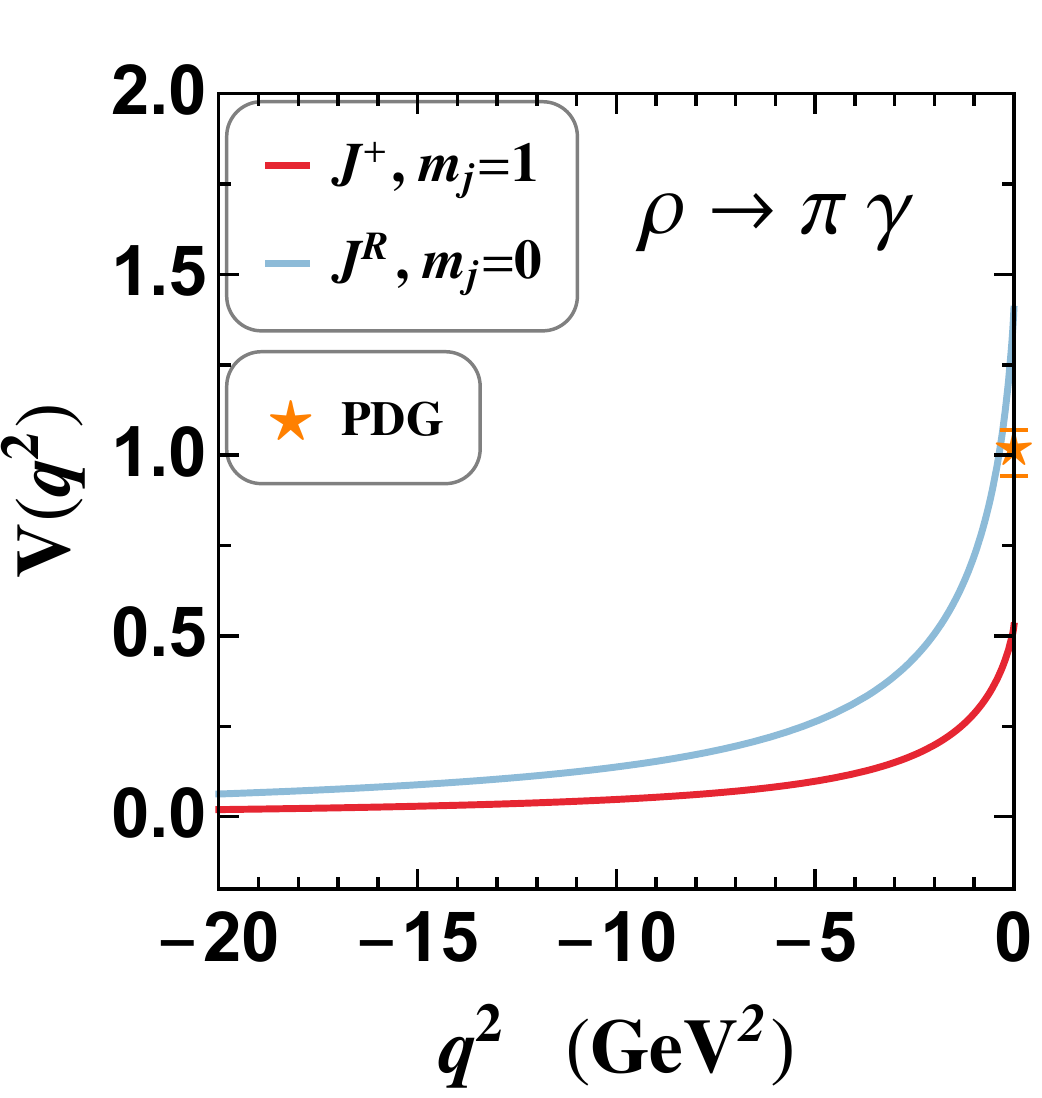}
\caption{Transition form factors calculated with 
  $J^R, m_j=0$
  and
  $J^+, m_j=1$ respectively. The LFWFs are calculated from the BLFQ approach, bottomonia and charmonia~\cite{Yang_run}
  ~and
light mesons~\cite{WQ_light}.
}
 \label{fig:R_pl_compare}
\end{figure}

  We compare $\hat V|_{J^{R/L}, m_j=0}$ and $\hat V|_{J^+, m_j=\pm 1}$ numerically in Fig.~\ref{fig:R_pl_compare}. The calculations are carried out with $z=0$, i.e. in the Drell-Yan frame.
  There are noticeable differences between the two for each of the three transitions. The preferred current, $\hat V|_{J^{R/L}, m_j=0}$, is closer to experiment values~\cite{PDG2018}.

\subsection{The preferred frame}
In general, the transition amplitude on the light front is given by the sum of the diagonal $n\to n$ and the off-diagonal $n+2\to n$ transitions ($n$ is the number of partons in the meson bound state). 
For calculations in the valence Fock sector, only the $2\to 2$ transition is considered whereas the non-valence contributions are absent. 
As from the LFWF representation of the hadron matrix elements in Eq.~\eqref{eq:jmuq}, the range of the longitudinal momentum fraction of the struck quark is $x'\in [z,1]$ before the transition and $x \in [0,1]$ after. The missing piece of $x'\in [0,z]$ corresponds to the hadron matrix elements of $n+2\to n$ involving higher Fock sectors. 
As a consequence, frames with minimal $z$ would increase the overlap region of the two wavefunctions in the valence contribution and suppress the non-valence contributions, and are thus preferred. They are: the Drell-Yan frame in the space-like region, and the longitudinal-II frame in the time-like region~\cite{Li:2019kpr}.

\begin{figure}[ht!]
        \centering
        \includegraphics[width=.3\textwidth]{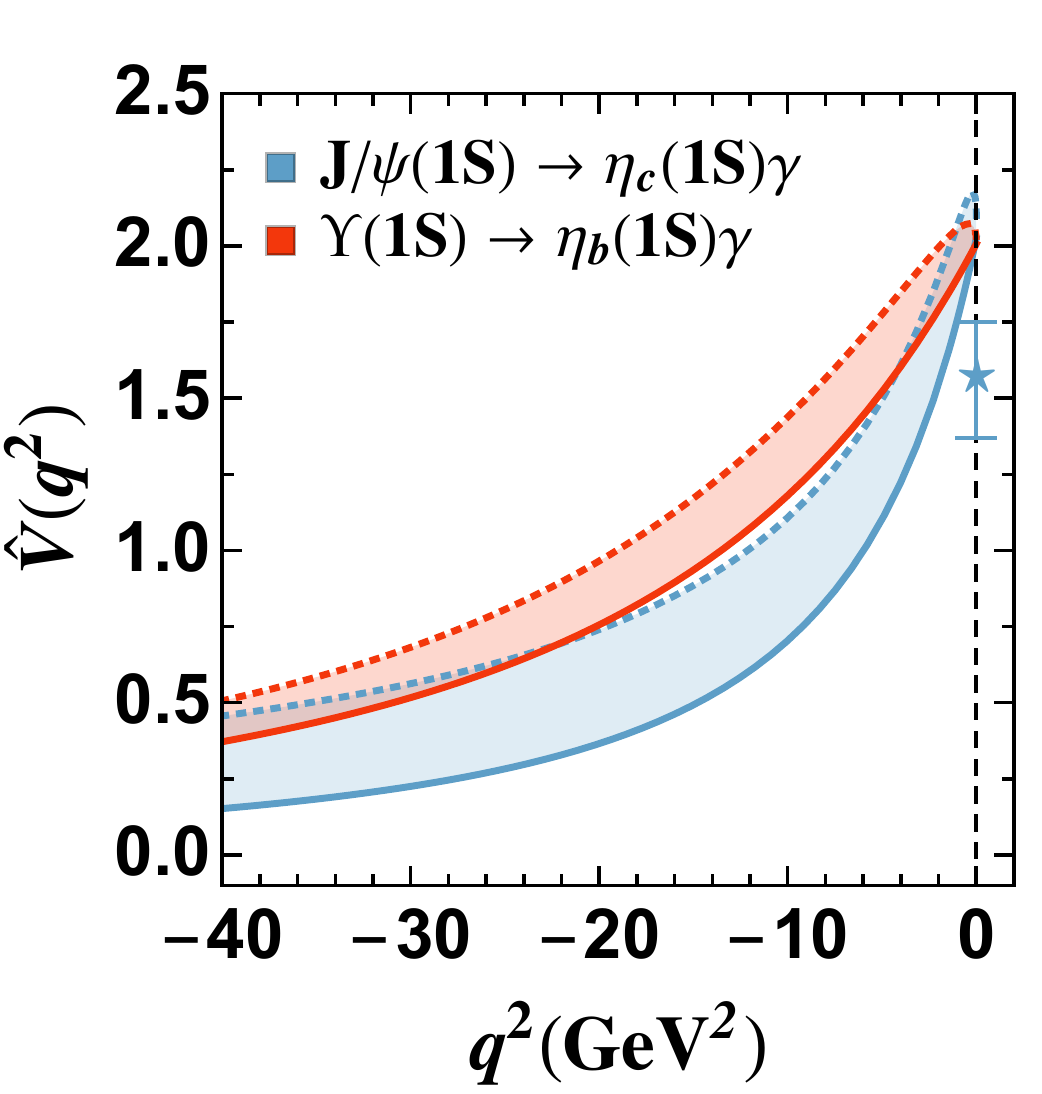}
        \includegraphics[width=.3\textwidth]{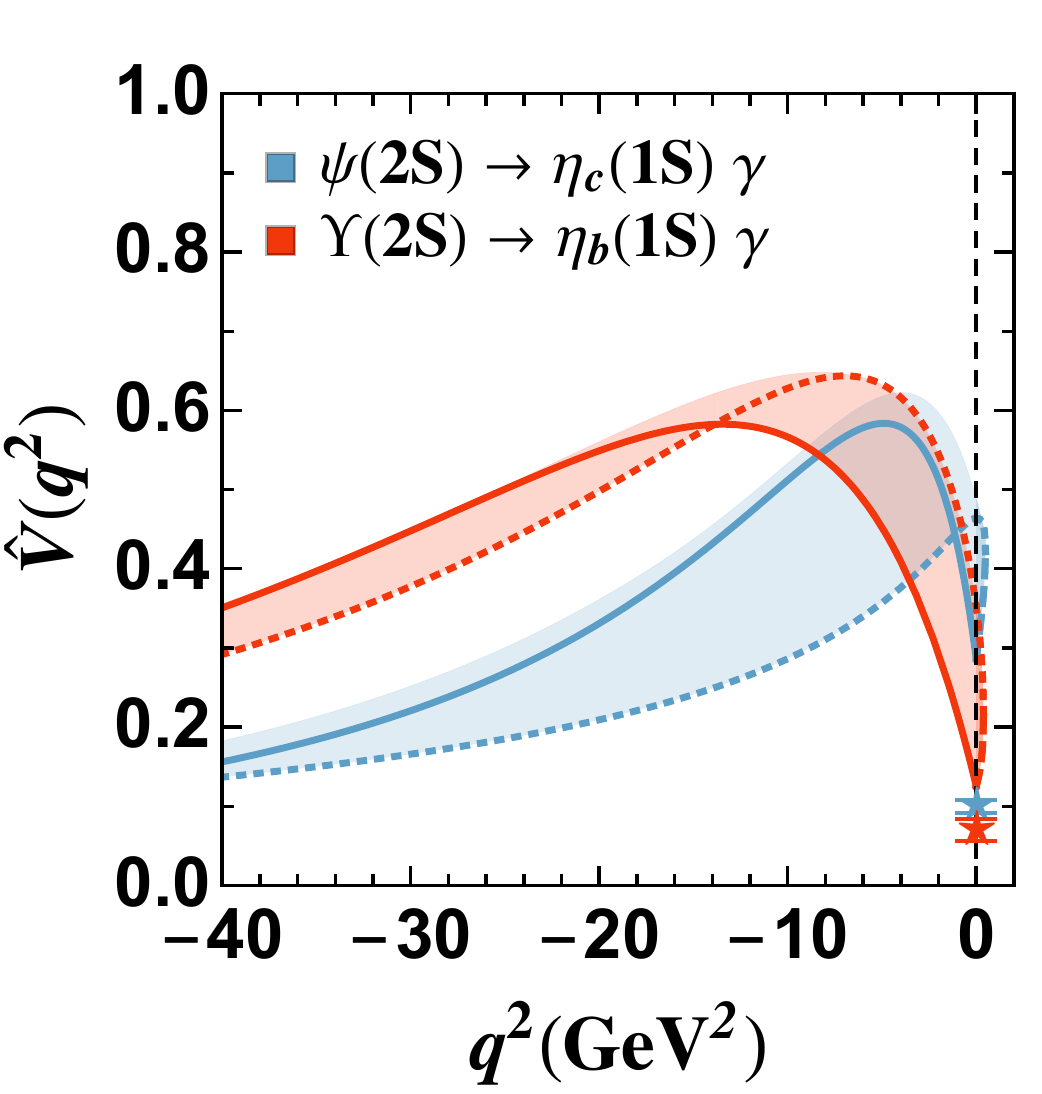}
        \includegraphics[width=.3\textwidth]{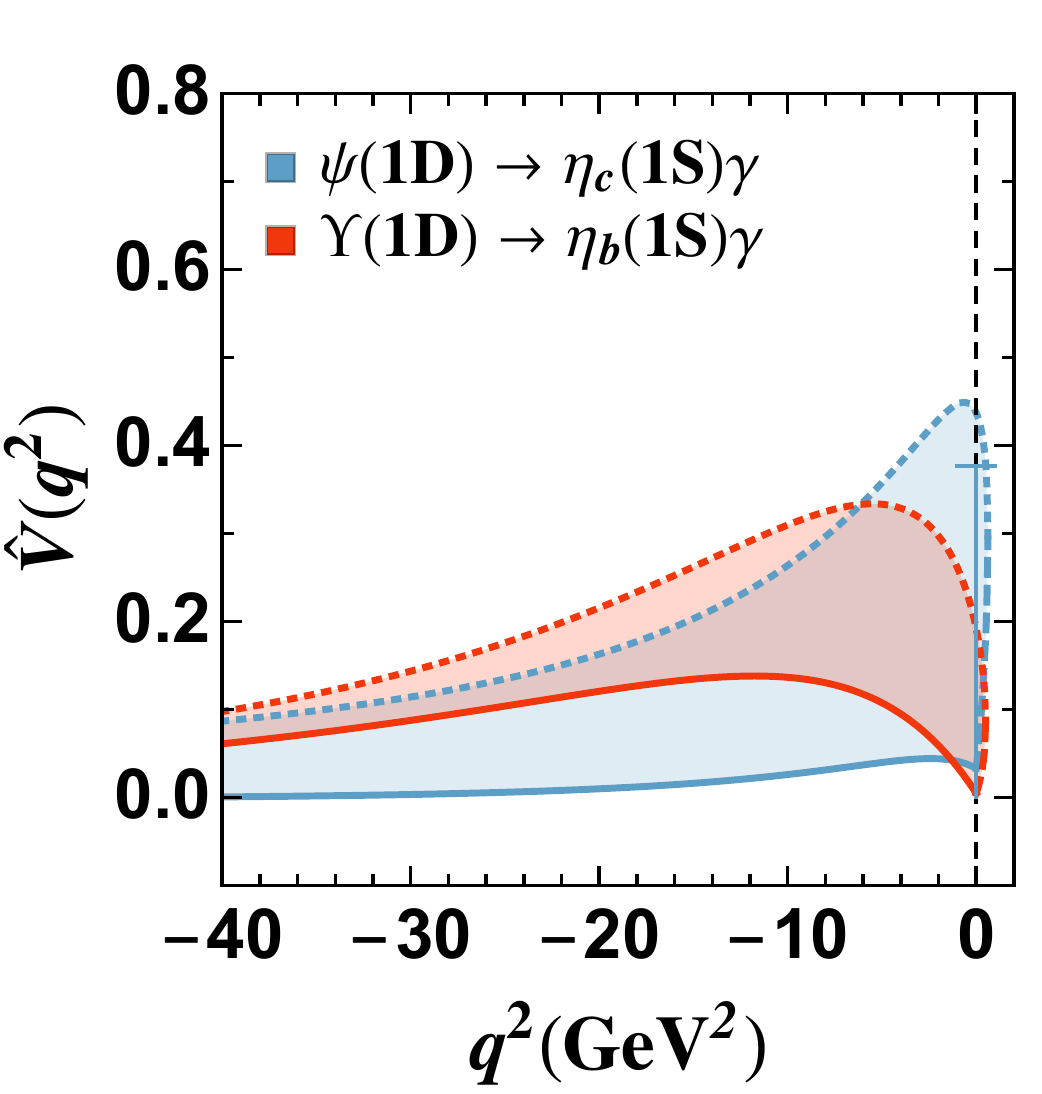}

        \includegraphics[width=.3\textwidth]{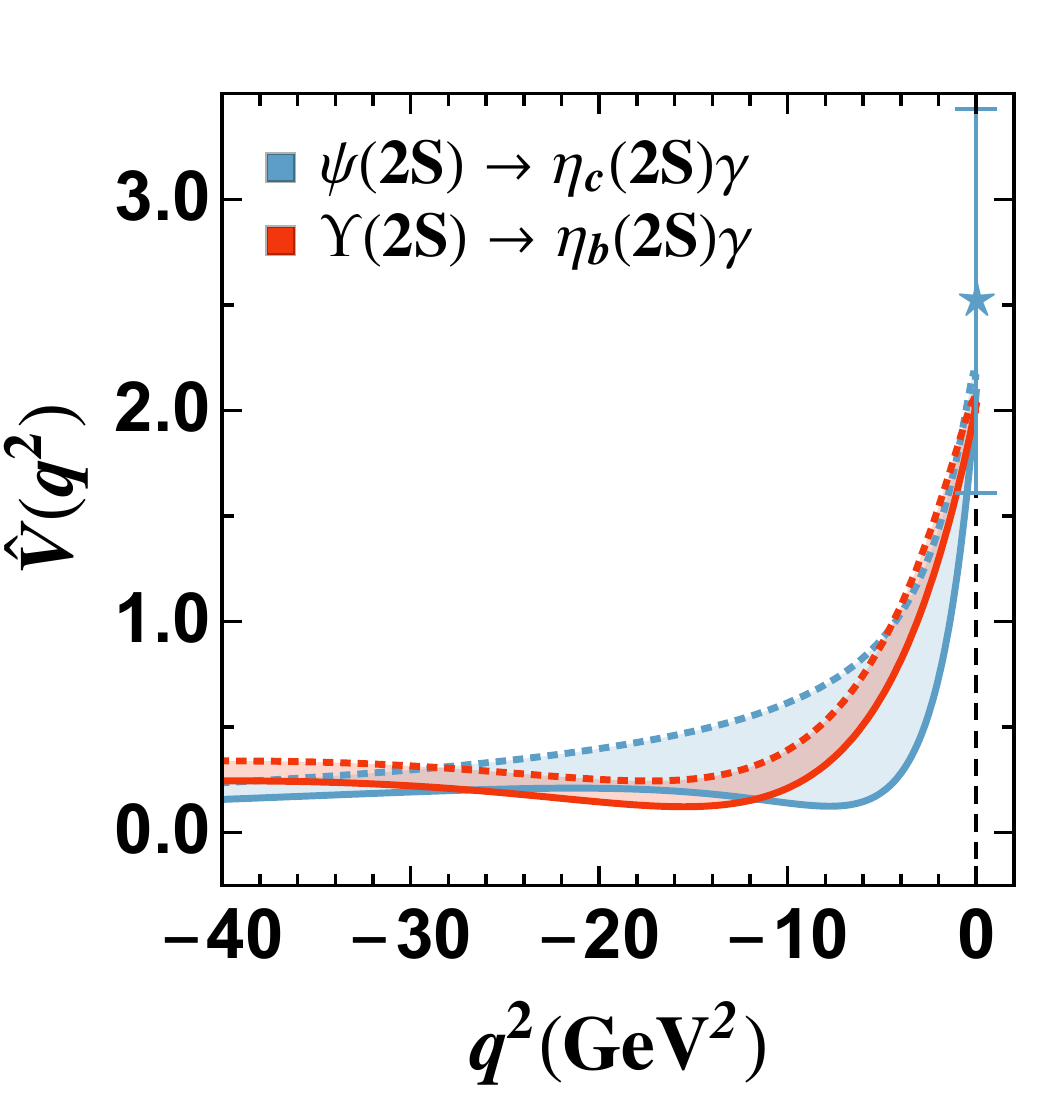}
        \includegraphics[width=.3\textwidth]{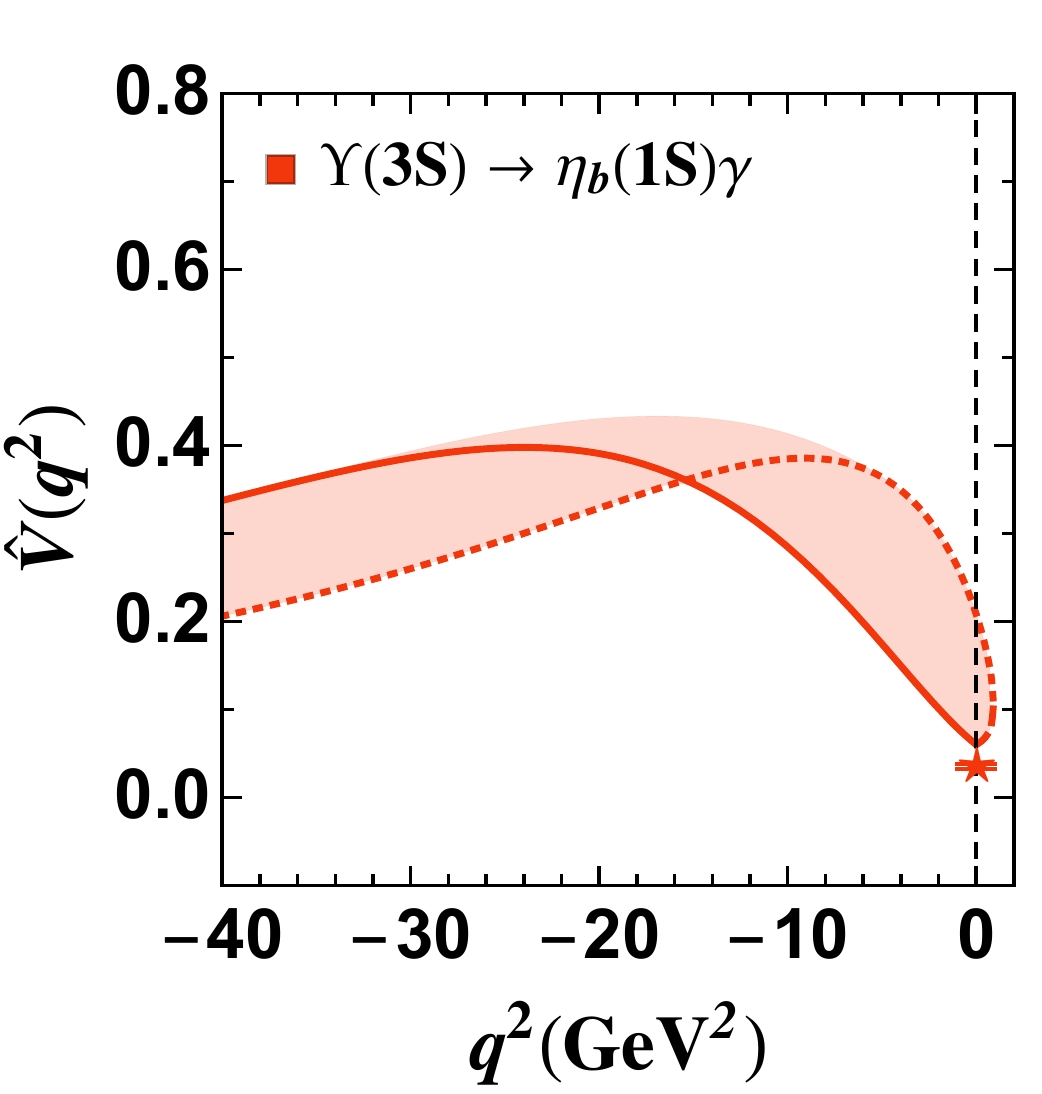}
        \includegraphics[width=.3\textwidth]{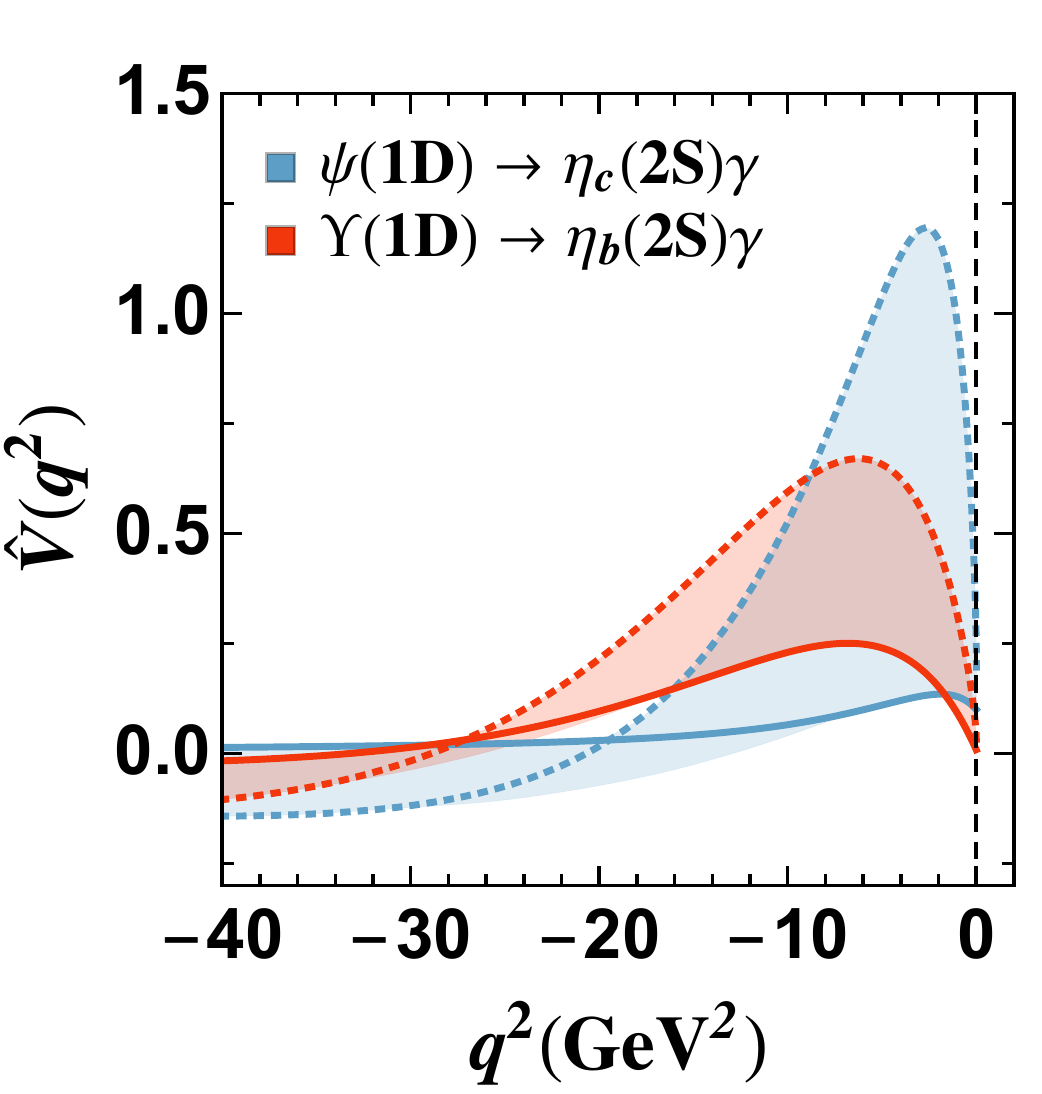}
      
        \caption{The transition form factor of the transition $\mathcal{V}\to\mathcal{P}\gamma$ of charmonia (blue curves/shades) and bottomonia (red curves/shades), calculated with LFWFs solved from the BLFQ approach~\cite{Yang_run}. 
        Meson masses are taken from experimental data~\cite{PDG2018} in defining the frames. The solid curves represent the Drell-Yan frame, the dotted and dashed lines represent the longitudinal I and II frames. The shaded areas represent the results from all other frames. The stars are values of $\hat V(0)$ converted from available decay widths in PDG~\cite{PDG2018}.
      }
       \label{fig:TFFccbb}
\end{figure}

We calculate the transition form factor through a dense sampling on the $(z,\vec \Delta_\perp)$ space with $\hat V|_{J^R,m_j=0}$ in Figure~\ref{fig:TFFccbb}.
The shaded areas represent the frame dependence. The suggested frames have a better agreement on $\hat V(0)$ with the experiment values indicated as stars~\cite{PDG2018}.

\section{Summary and Outlook}
We studied the M1 transition form factor on the light front with different current components and different frames. 
For calculations in the valence Fock sector, we suggest using the transverse current with $m_j=0$ state of the vector meson and frames with minimal longitudinal momentum fraction.
In future work, we would like to include higher Fock sectors in solving the meson bound states and further investigate the rotational symmetry and its roles in radiative transitions.

\section{Acknowledgments}
I thank the International Light Cone Advisory Committee for awarding me the McCartor Fellowship. I acknowledge valuable discussions with Y. Li, P. Maris, J. P. Vary, S. Jia, T. Frederico, S. Tang, W. Qian, H.-M. Choi, C.-R. Ji and S. D. G$\l{}$azek. This research was supported in part by the U.S. Department of Energy (US-DOE) under Grant No. DE-FG02-87ER40371,
and used resources of the National Energy Research Scientific Computing Center, a US-DOE Office of Science User Facility operated under Contract No.DE-AC02-05CH11231.
\bibliographystyle{JHEP} 
\bibliography{TFF_Dalitz}


\end{document}